\documentclass[aps,prb,showpacs,twocolumn,groupedaddress]{revtex4-1}

\usepackage{graphicx,graphics,color,epsfig}
\usepackage{amsmath}
\usepackage{amssymb}
\usepackage{amsfonts}
\usepackage{amsfonts}
\usepackage{epstopdf}
\usepackage{bm}
\usepackage{times,xspace}
\usepackage{color}

\begin{document}

\preprint{PRL}

\title{{\em Q}=0 collective modes originating from the low-lying Hg-O band in the superconducting HgBa$_2$CuO$_{4+\delta}$}

\author{Tanmoy Das}
\affiliation{Theoretical Division, Los Alamos National Laboratory, Los Alamos, New Mexico 87545 USA}

\date{\today}

\begin{abstract}
Motivated by the recent discovery of two ${\bm Q}\sim0$ collective modes [Y. Li {\it et al.}, Nature {\bf 468}, 283 (2010); {\it ibid} Nat. Phys. {\bf 8}, 404 (2012)] in the single-layer HgBa$_2$CuO$_{4+\delta}$, which are often taken as evidence of the orbital current origin of a pseudogap, we examine an alternative and assumption-free scenario constrained by first-principle calculations. We find that in addition to the common CuO$_2$ band, a hybridized Hg-O state is present in the vicinity of the Fermi level, and that it contributes to the low-energy ground state of this system. We calculate the spin-excitation spectrum based on the random-phase-approximation in the superconducting state using a two band model and show that a collective mode in the multi-orbital channel arises at ${\bm Q}=0$. This mode splits in energy, yet remains at ${\bm Q}\sim0$ as the pseudogap develops breaking both translational and time-reversal symmetries in the spin-channel, and thus gap out the electrnic states. The observations of the dynamical mode and static moment in the pseudogap state are in good accord with experimental observations. A verifiable prediction of this proposal is the development of magnetic order also in the Hg-O layer. Detection of Hg-O band via optical study, or magnetic moment in the Hg-O layer will be tests of this calculation.
\end{abstract}

\maketitle

Unraveling the nature and mechanism of the pseudogap has remained a steady theme of the research of cuprates,\cite{PGcuprate} and has recently been extended to pnictide\cite{PGpnictide} and heavy fermion systems.\cite{sasha} Within cuprates, the evidence of pseudogaps has emerged to be highly contradictory in different experimental probes as well as in different materials. While the original idea of the `pseudogap' stems from the observations of anomalous (non mean-field like) temperature dependencies in various bulk measurements,\cite{pgbulk} and incoherent spectral weight properties\cite{pgarpes} in spectroscopies that set in below a characteristic temperature $T^*$, recent magneto-transport measurements have indicated the presence of some form of density wave origin of it.\cite{pgmagtransport}. Focusing on the broken symmetry state of the pseudogap from various inelastic neutron scattering (INS) data, its nature apparently seems to be strongly material dependent within the cuprate family as follows. (1) Electron doped cuprates consistently demonstrate the presence of a commensurate antiferromagnetic order up to the superconducting (SC) region.\cite{GrevenNCCO} (2) In single layer hole doped La$_{2-x}$(Ba/Sr)$_x$CuO$_4$ (LB/SCO), the commensurate order is observed to become incommensurate with doping (with a so-called `hour-glass' dispersion in the spin-excitation spectrum), and rotates along the Cu-O bond direction as the SC dome develops.\cite{Yamadaplot} Such phenomena have been taken as evidence of the `stripe' order origin of the pseudogap in these systems.\cite{tranquada} (3) In YBa$_2$Cu$_3$O$_{6+\delta}$ (YBCO), the incommensurate spin-excitation further exhibits an in-plane anisotropy, having stronger intensity along the $a$-bond direction than the $b$-direction, a fact which has been interpreted as the emergence of electronic nematic order.\cite{Hinkov} (4) In addition to the `hour-glass' pattern, which is a trademark of all these hole doped systems, a ${\bm Q}\sim 0$ mode is observed in single layer HgBa$_2$CuO$_{4+\delta}$ (Hg1201),\cite{Grevenonemode} which further splits in energy.\cite{Greventwomode} Such an Ising-like mode can be expected within a circulating orbital current model.\cite{Varma}

Understanding such diverse material dependence of the spin-excitation properties in cuprates within a single model has remained a challenge. However, some efforts to obtain a unified model interpretation should be mentioned. The extension of the model of coexistence of an (either long- or short-ranged) spin-density wave (SDW) and $d$-wave superconductivity for electron doped cuprates to the hole doped side is one of these successful approaches.\cite{Dastwogap,Dasresonance} Within this model, it has been shown that the commensurate mode is lifted to higher energy by the SC gap with a downward dispersion reaching to zero energy due to the nodal $d$-wave momentum $({\bm k})$ dependence of the SC gap.\cite{Dasresonance,spinextheory} In other words, the spin-excitation manifests itself as static incommensurate peaks via coupling to the nodal $d$-wave SC gap. This model reproduces the `hour-glass' pattern, a 45$^o$ rotation of the incommensurate spectra in all these systems. Furthermore, the same model has been applied to quantitatively explain the additional in-plane anisotropy in YBCO by incorporating the inter-layer coupling between the CuO$_2$ plane and the metallic {\it unidirectional} CuO chain plane which is present at all finite dopings.\cite{Daschain} This calculation does not require any `spontaneous' nematic order. In the present paper, we extend the model to the Hg1201 system to explain the coexistence of both the `hour-glass' pattern and the ${\bm Q}\sim0$ modes.

\begin{figure}[top]
\hspace{-0cm}
\rotatebox{0}{\scalebox{.46}{\includegraphics{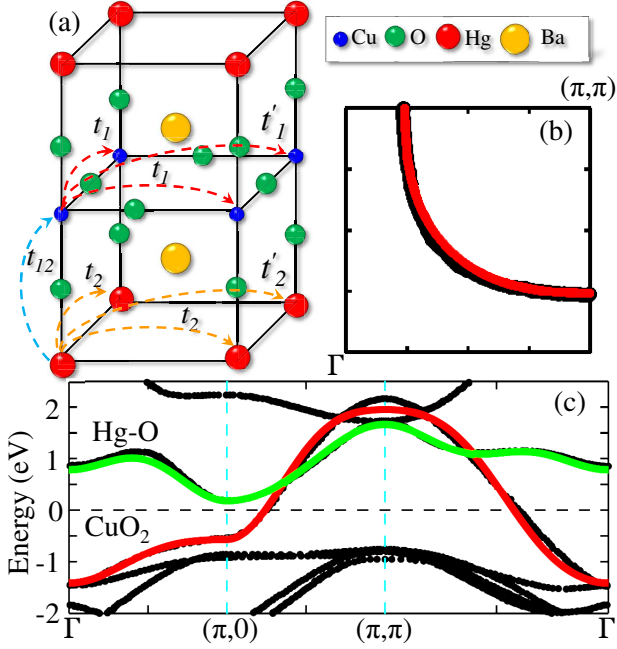}}}
\caption{(Color online) (a) Crystal structure and tight-binding hopping parameters of the Hg1201 system. (b) Non-interacting tight-binding FS (red solid line) plotted on top of the first-principles result.\cite{LDAold,LDAAoki,LDAMoreira} (c) Corresponding dispersion plotted along high-symmetry directions in solid colors and compared with {\it ab-initio} dispersions.}
\label{fig1}
\end{figure}

Looking into the first-principles band structure,\cite{LDAold,LDAAoki,LDAMoreira} we find that there exists a Hg-O band which lies so close to the Fermi level ($E_F$) that it can contribute to the low-energy physics in this compound. Constrained by this band-structure property, we present an assumption-free two band tight-binding model including the CuO$_2$ plane and Hg-atomic plane, see Fig.~\ref{fig1}(a). We compute the spin excitation spectrum to show that due to the interaction between these two plane states, a ${\bm Q}\sim0$ collective mode develops in the multi-orbital spin channel. We further demonstrate that when a SDW order or any other similar translational symmetry breaking order sets in the CuO$_2$ plane in the pseudogap regime, it induces a magnetic moment to the Hg-O state which splits the ${\bm Q}\sim0$ mode in energy, but not in momentum. The observations of the dynamical mode and associated static moment in the pseudogap state are in good accord with experimental observations.\cite{Grevenonemode,Greventwomode,morderHg}

{\it Tight-binding Hamiltonian:-} Unlike in other cuprates where the Cu-O-O layer(s) dominate the most interesting low-energy electronic states, in the Hg-based cuprates the hybridization of the reservoir atom Hg with the apical oxygens and Ba is strong, making the Hg-O derived band cross or stay close to $E_F$. Both x-ray photoemission data\cite{PES} and first-principles\cite{LDAold,LDAAoki,LDAMoreira} calculations have demonstrated the presence of the bottom of the Hg-O band lying as close as 0.09-0.13~eV above $E_F$ at the `X' point which moves and even crosses $E_F$ with doping, interaction, or with an increasing number of CuO$_2$ layers. The first-principles calculations also indicate that individual CuO$_2$ and Hg-O bands are highly two-dimensional, and the inter-layer coupling between them is prominent. Based on these considerations, we derive a realistic two-band tight-binding model for Hg1201, in which the parameters are deduced by fitting to first-principle dispersion, without any adjustments. The obtained non-interacting Hamiltonian is
\begin{eqnarray}
H_{\bm{k}}=
\left(
\begin{array}{cc}\
\xi_{1{\bm k}}  & \xi_{12{\bm k}} \\
\xi_{12{\bm k}} & \xi_{2{\bm k}}
\end{array}
\right).\label{Ham0}
\end{eqnarray}
The dispersions $\xi_{1{\bm k}}$ and $\xi_{2{\bm k}}$ are for the CuO$_2$ and Hg-O states, respectively, where $\xi_{12{\bm k}}$ is the inter-layer coupling between them, see Fig.~\ref{fig1}(a), in which $\xi_{i{\bm k}}=-2t_i(\phi_x+\phi_y)-4t_i^{\prime}\phi_x \phi_y -2t_i^{\prime\prime}(\phi_{2x}+\phi_{2y})-4t_i^{\prime\prime\prime}(\phi_{2x}\phi_y+\phi_{2y}\phi_{x})-\mu_i$, and $\xi_{12{\bm k}} =-2t_{12}\phi_{z/2}$, where $\phi_{\alpha x/y/z}=\cos{(\alpha k_{x/y/z})}$. The corresponding values of the tight-binding parameters $t_i$ and chemical potentials $\mu_i$ are given in Ref.~\onlinecite{tbparam}.

The tight-binding fittings to the first-principles dispersion and Fermi surface (FS) are given in Figs.~\ref{fig1}(b) and \ref{fig1}(c). The CuO$_2$ antibonding state is clearly visible , and it only constitutes the FS. An essential difference between the CuO$_2$ and Hg-O layers is that despite the absence of O-atoms in the Hg-plane, see Fig.~\ref{fig1}(a), first-principles calculations exhibit that the latter state (the green line above $E_F$) originates from the hybridization between the Hg and O atoms. This crucial information emphasizes that the inter-layer tunneling matrix-element $t_{12}$ is important and cannot be neglected for any realistic computation. Based on this observation, it is justifiable to assume that the inter-layer interaction, $V$, is also strong, and it is the prime term for the development of the ${\bm Q}\sim 0$ mode in our model.

\begin{figure}[top]
\hspace{-0cm}
\rotatebox{0}{\scalebox{.46}{\includegraphics{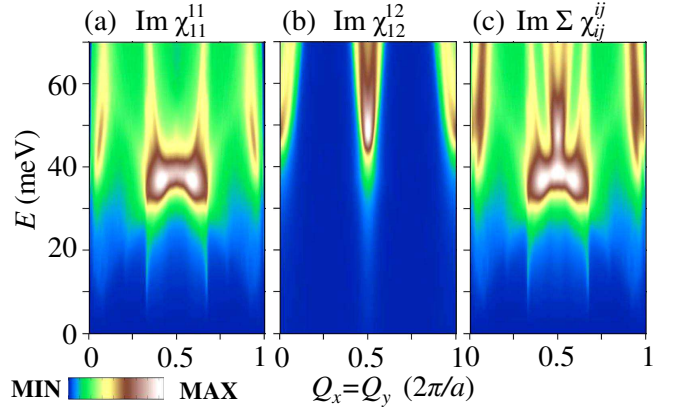}}}
\caption{(Color online) Imaginary part of the BCS-RPA susceptibility in the paramagnetic state plotted along the diagonal direction as a function of excitation energy. (a) The intra-atomic component for the CuO$_2$ band reveals strong peak in the ${\bm Q}_2\sim(\pi,\pi)$ region. Although in the bare level, this component does not contain any other intensity except at ${\bm Q}_2$, its RPA value involves weak features near ${\bm Q}_1\sim 0$ due to mixing with other terms within the tensor form of the RPA denominator (discussed in the text). (b) The inter-atomic RPA susceptibility between the CuO$_2$  and Hg-O bands shows intensity at  ${\bm Q}_1$. The zone periodicity of the intensity between ${\bm Q}_1$ and ${\bm Q}_2$ even in the paramagnetic state strongly suggests that the ${\bm Q}_1$ mode is unstable to a (fluctuating) magnetic ground state with a modulation of ${\bm Q}_2$. (c) The total RPA susceptibility yields strong intensity at both ${\bm Q}$ vectors of different origins, in agreement with experimental data.\cite{Grevenonemode}}
\label{fig2}
\end{figure}

Before including magnetic order, we first study the physical origin of the ${\bm Q}_1\sim 0$ and ${\bm Q}_2\sim(\pi,\pi)$ modes in the paramagnetic state. The widely believed origin of the spin-resonance mode at ${\bm Q}_2$ is due to the the sign-reversal of the $d-$wave pairing at the magnetic `hot-spot' on the FS of the CuO$_2$ state.\cite{spinextheory,Dasresonance} Here we show that the ${\bm Q}_1$ mode is also a collective phenomenon of different origin that develops in the particle-hole continuum of the the inter-orbital channel within the random-phase approximation (RPA). Due to the lack of a FS in the Hg-O band, we assume this band to be non-SC in this single layer Hg1201 case. Including all these realistic effects, we evaluate the orbital dependent non-interacting BCS susceptibility as
\begin{eqnarray}
&&(\chi_0)^{ii^{\prime}}_{j{j^{\prime}}}({\bm q},p_m) = -\frac{1}{2}\sum_{{\bm k},n,\nu\nu^{\prime}} M^{\nu\nu^{\prime}}_{ii^{\prime}j^{\prime}j}({\bm k},{\bm q})\nonumber\\
&&\times \left[G^{\nu}_{\bm k}(\omega_n)G^{\nu^{\prime}}_{{\bm k}+{\bm q}}(\omega_n+p_m)-F^{\nu}_{\bm k}(\omega_n)F^{\nu^{\prime}}_{-{\bm k}-{\bm q}}(\omega_n+p_m)\right].\nonumber\\
\label{eq:chi0}
\end{eqnarray}
Here $G^{\nu}$ and $F^{\nu}$ are the normal and anomalous Green's functions for the quasiparticle band $E^{\nu}_{\bm k}=\sqrt{(\epsilon^{\nu}_{\bm k})^2+(\Delta^{\nu}_{\bm k})^2}$ where $\epsilon^{\nu}_{\bm k}$ are the eigenvalues of the Hamiltonian given in Eq.~\ref{Ham0}. If the corresponding eigenstates are denoted by $\psi^{\nu}_i({\bm k})$ for the $i^{th}$ orbital, then the orbital overlap matrix-element can be written as $M_{ii^{\prime}j^{\prime}j}({\bm k},{\bm q})=\psi_i^{\nu\dag}({\bm k})\psi_{i^{\prime}}^{\nu^{\prime}}({\bm k}+{\bm q})\psi_{j^{\prime}}^{\nu^{\prime}\dag}({\bm k}+{\bm q})\psi_{j}^{\nu}({\bm k})$. $\omega_n$ and $p_m$ are the fermionic and bosonic Matsubara frequencies, respectively. The SC gap functions are taken as $\Delta_{\bm k}^1 =\Delta_0(\phi_{x}-\phi_{y})$ and $\Delta_{\bm k}^2=0$ for the CuO$_2$ and Hg-O bands, respectively, with $\Delta_0$=31~meV. Finally, we employ a RPA correction to obtain the many-body orbital spin susceptibility  as $\tilde{\chi}({\bm q},\omega)=\tilde{\chi_0}({\bm q},\omega)[{\bf 1}-\tilde{U}_s\tilde{\chi_0}({\bm q},\omega)]^{-1}$, where $\tilde{U}_s$ is the interaction tensor in the transverse spin-flip channel defined in the orbital basis (see Ref.~\onlinecite{RPAU}).

The individual and the total components of the RPA-BCS susceptibility are presented in the SC state in Fig.~\ref{fig2}. The imaginary part of the intra-orbital component $\chi_{11}^{11}$ for the CuO$_2$ state clearly exhibits the lower-branch of the so-called `hour-glass' pattern at  ${\bm Q}_2$, which is the trademark feature of the $d$-wave SC gap as ubiquitously measured by INS in all hole-doped cuprates,\cite{tranquada,Hinkov,Bi2212} including the present Hg1201 system.\cite{Greventwomode} In the inter-orbital channel $\chi_{12}^{12}$, a strong resonance mode appears at ${\bm Q}_1\sim0$, with a somewhat upward dispersing branch with vanishing intensity. Tracking down to the band-structure details in Fig.~\ref{fig1}(c), we reveal that this mode originates from the direct excitation gap in the van-Hove singularity region (the RPA correction shifts the mode energy to a lower value). It is interesting to note that even in the absence of any magnetic order, the ${\bm Q}_1$ mode reveals zone periodicity at $(\pi,\pi)$, suggesting that the Hg-atoms tend to magnetically order by the same wave vector. Since a ${\bm Q}_1\sim0$ mode develops in the paramagnetic ground state, it is expected to survive in the overdoped region. In this context, we recall a recent observation of an unusual Raman mode\cite{Raman} in overdoped Hg1201 which can be taken as the persistence of ${\bm Q}_1\sim0$ above the pseudogap region. Optical absorption spectroscopy which measures the direct gap can also be used to test our proposal.

\begin{figure}[top]
\hspace{-0cm}
\rotatebox{0}{\scalebox{.46}{\includegraphics{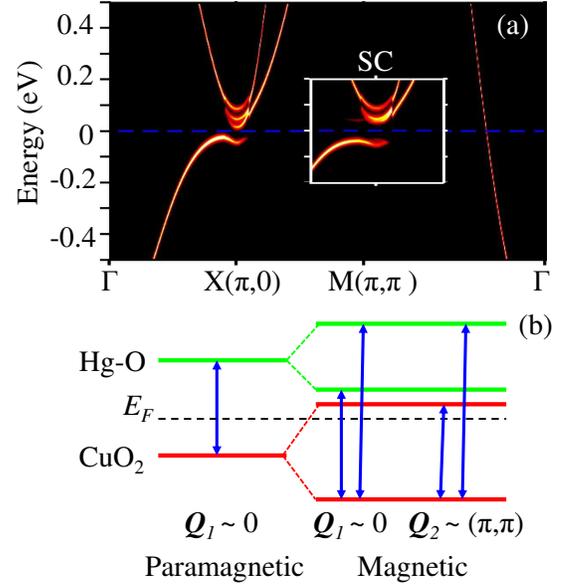}}}
\caption{(Color online) (a) Single particle spectral weight map in the SDW state. The gap opening occurs in CuO$_2$ at $E_F$ at the antinodal region, while the smaller magnetic gap in the Hg-O state commences above $E_F$. {\it Inset:} As superconductivity is turned on in the CuO$_2$ state, the corresponding gap size increases, and for these particular gap values, the upper CuO$_2$ magnetic band and lower Hg-O magnetic band becomes close to each other. (b) The details of the energy level and the corresponding spin excitation transitions are illustrated in the paramagnetic and magnetic states. At ${\bm Q}_2$, the magnetic transition across $E_F$ also involves a momentum transfer, as expected, which is not explicitly illustrated in this schematic diagram.}
\label{fig3}
\end{figure}

{\it Magnetic ground state:-} Next we focus on how a magnetic order can split the ${\bm Q}_1\sim0$ mode. Earlier nuclear magnetic resonance (NMR) data on the same sample have demonstrated the opening of the pseudogap at $T^*$ as in the INS measurements.\cite{NMR} Recent neutron diffraction data on the same sample\cite{morderHg} as well as in YBCO\cite{morderYBCO1,morderYBCO2} establish that there exists a static magnetic order which vanishes above $T^*$, in addition to the dynamical mode that saturates above $T^*$. Relating the static magnetic order to that that which renders the FS reconstruction, we work in an (in-plane) double unit cell magnetic Brillouin zone connected by the commensurate SDW wave vector ${\bm Q}_2=(\pi,\pi)$. If no other symmetry is broken, the new magnetic zone is the same for both the CuO$_2$ plane as well as the Hg-plane in the tetragonal lattice, and thus, a magnetic moment is induced to the latter state. Using the standard Nambu notation, we define the two atomic eigenstates in the magnetic zone as $\Psi_{\bm k}^{\dag}=\left[ c_{1{\bm k}\uparrow}^{\dag},~ c_{2{\bm k}\uparrow}^{\dag},~ c_{1({\bm k}+{\bm Q}_2)\downarrow}^{\dag},~c_{2({\bm k}+{\bm Q}_2)\downarrow}^{\dag}\right]$, where $c_{i{\bm k}\sigma}^{\dag}~(c_{i{\bm k}\sigma})$ creates (annihilates) an electron with momentum ${\bm k}$ and spin $\sigma$ on the $i^{th}$ atom. In this notation the Hamiltonian presented in Eq.~\ref{Ham0} becomes a 4$\times$4 matrix $H = \sum_{\bm k}\Psi_{\bm k}^{\dag}H_{\bm k}\Psi_{\bm k}$:
\begin{eqnarray}
H_{\bm{k}}=
\left(
\begin{array}{cccc}\
\xi_{1{\bm k}} & \xi_{12{\bm k}} & -U_1m_1 & 0 \\
\xi_{12{\bm k}} & \xi_{2{\bm k}} &  0 & -U_2m_2\\
-U_1m_1 & 0 & \xi_{1({\bm k}+{\bm Q}_2)} & \xi_{12({\bm k}+{\bm Q}_2)}\\
0 & -U_2m_2 &\xi_{12({\bm k}+{\bm Q}_2)} & \xi_{2({\bm k}+{\bm Q}_2)}\\
\end{array}
\right).\label{Hamk}
\end{eqnarray}
\begin{figure}[top]
\hspace{-0cm}
\rotatebox{0}{\scalebox{.7}{\includegraphics{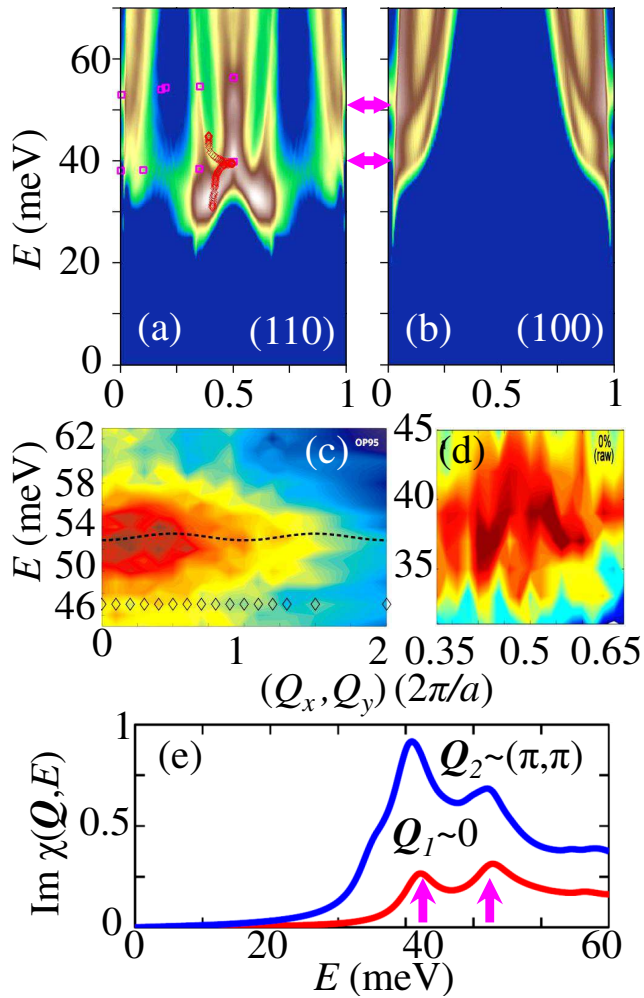}}}
\caption{(Color online) (a), (b) Theoretical spin-excitation spectra in the coexistence of the SDW and SC ground state, plotted along the diagonal, and the (100) directions respectively. The symbols give the corresponding experimental data for Hg1201 near optimal doping.\cite{Grevenonemode} The splitting of the ${\bm Q}_1$ mode in the pseudogap state is more evident in (b). (c), (d) Intensity plots of the same experimental results near ${\bm Q}_1$ and ${\bm Q}_2$, respectively. It is evident from the experimental data that despite the presence of a tail of the weakly dispersing branch in (c), the strong intensity is concentrated near ${\bm Q}_1$ as in our theoretical spectra in (b). (e) The same theoretical result plotted at ${\bm Q}_1$ and ${\bm Q}_2$, demonstrating the relative intensities of the corresponding peaks. Here the two split modes at ${\bm Q}_1$ with equivalent intensities are clearly visible.}
\label{fig4}
\end{figure}

Here the order parameters $m_i$ represent the staggered magnetic moments, which in the mean-field level are evaluated self-consistently as $m_i=\sum_{{\bm k}\sigma}\sigma\left\langle c_{i({\bm k}+{\bm Q}_2)\sigma}^{\dag}c_{i{\bm k}\bar{\sigma}}\right\rangle =\sum_{j\ne i,{\bm k}\sigma}\int_{\infty}^{\infty}\frac{d\omega}{2\pi}\sigma A_{ij}({\bm k},\sigma,\omega)f(\omega)$. Here the spin-resolved spectral function $A_{ij}({\bm k},\sigma,\omega)=-{\rm Im}G_{ij}({\bm k},\sigma,\omega)/\pi$, where $f$ is the Fermi function, and $\sigma=\bar{\sigma}=\pm$. Self-consistent values of the magnetic moments are 0.05$\mu_B$ and 0.04$\mu_B$ in the CuO$_2$ and Hg-O bands, respectively, and their total value is close to the experimental result of 0.1$\mu_B$ for the Hg1201\cite{morderHg} and YBCO\cite{morderYBCO1,morderYBCO2} systems.

The single-particle spectral function $\sum_{i\sigma} A_{ii}({\bm k},\sigma,\omega)$, plotted along the high-symmetry directions in Fig.~\ref{fig3}(a) exposes the nature of gap opening in both bands. The direct gap values between the spin-split states is determined by $Um_i$ which increases to $\sqrt{(Um_1)^2+(\Delta^1_{\bm k})^2}$ in the SC state in the CuO$_2$ layer. (There is no SC gap in the Hg-O state). In the magnetic state, $G$ and $F$ also become 4$\times$4 tensors, and with this modification, the calculations of the spin resonance spectrum remain the same as before. Several low-energy magnetic transition channels in the particle-hole continuum become active in this case as illustrated in Fig.~\ref{fig3}(b). Earlier calculations have shown that the SDW order in the CuO$_2$ state gives rise to an upward dispersion centering on ${\bm Q}_2$ which meets the downward dispersion of the SC origin at  ${\bm Q}_2$ to create the so-called `hour-glass' phenomenon.\cite{Dasresonance} Such a pattern is reproduced here in the CuO$_2$ state as shown in Fig.~\ref{fig4}(a). Due to the lack of a FS in the Hg-O state, such upward dispersion is absent in this band. However, in multi-layered HgBa$_2$Ca$_{n-1}$Cu$_n$O$_{2n+2}$ (with $n>1$) the Hg-O band crosses below $E_F$, and based on the same theoretical argument\cite{Dasresonance} an additional upward branch can be expected there.

In Fig.~\ref{fig4} we give details of the evolution of two ${\bm Q}_1\sim 0$ modes in the magnetic state. As illustrated in Fig.~\ref{fig3}(b), the excitation from the lower-magnetic band of the CuO$_2$ state to the two split Hg-O bands above $E_F$ creates two spin resonance modes.  This energy splitting is evident in the spin-excitation spectrum plotted along the diagonal and (100)-directions in Figs.~\ref{fig4}(a), and \ref{fig4}(b), respectively. We include the experimental data (symbols) for the optimally doped Hg1201 sample for comparison.\cite{Greventwomode} The ${\bm Q}_1\sim 0$ mode splits by about 20~meV and lies at 45 and 55~meV, in accord with experimental values. Some discrepancies are clearly visible. The intensity plot of the  experimental data [reproduced from Ref.~\onlinecite{Greventwomode} in Figs.~\ref{fig4}(c), and \ref{fig4}(d)] reveals a weak dispersion mode with its intensity sharply vanishing away from the ${\bm Q}=0$ momentum. We find that the tail of the intensity in the theoretical curve disperses strongly to higher energy than its experimental counterpart. On the other hand, experimental data also demonstrated  that while the intensity of the ${\bm Q}_2$ mode vanishes at $T^*$, the same at ${\bm Q}_1$ presumably merges to the background. This supports our postulate that the latter is not directly related to the pseudogap physics, but arise in the particle-hole continuum of the paramagnetic ground state.

\section{Conclusion}

In summary, we provide a different and realistic explanation to the experimental observations of the Ising-like spin excitation modes in Hg1201 based on the presence of additional Hg-O states close to $E_F$ as established by first-principles and photoemission electronic structure considerations. The higher energy ${\bm Q}_1\sim 0$ mode is fairly doping independent, and based on our calculation we predict that it should survive to overdoping. On the other hand the second mode appears as a result of the proximity induced magnetic order in the Hg-atoms and hence it is doping and temperature dependent. These facts are in accord with experimental data. The similar observation of the ${\bm Q}_1\sim 0$ mode in the YBCO sample\cite{morderYBCO1,morderYBCO2} can also be explained within the same framework due to the presence of a metallic chain state in this sample.\cite{Daschain} Furthermore, in double layered Bi$_2$Sr$_2$CaCu$_2$O$_{8+\delta}$  (Bi2212), first-principles calculations\cite{Bi2212} find that the Bi-O band lies close to the Fermi level or may even cross it near the same $(\pi,0)$ point and thus can give rise to a similar ${\bm Q}\sim0$ magnetic mode, but its energy scale is yet to be explored. However, despite extensive neutron studies in other hole-doped cuprates, any signature of this ${\bm Q}\sim 0$ mode has not yet been reported supporting our theory. The gap opening in the Hg-O state can be tested by scanning tunneling microscopy, optical and x-ray absorption spectroscopies, whereas the prediction of magnetic order in Hg-O layer can be verified via NMR, neutron measurements.

\begin{acknowledgments}
The author acknowledges useful discussions with A. V. Balatsky, M. J. Graf, M. Greven, R. S. Markiewicz. The work is supported by the U.S. DOE through the Office of Science (BES) and the LDRD Program and benefited by NERSC computing allocation.
\end{acknowledgments}


\begin{thebibliography}{10}
\bibitem{tbparam}The values of the tight-binding parameters are evaluated by fitting to the LDA dispersion, as given in Fig.~\ref{fig1}, and the obtained values are $(t_1,~t^{\prime}_1,~t^{\prime\prime}_1,~t^{\prime\prime\prime}_1,~\mu_1)$=(0.46,~-0.105,~0.08,~-0.02,~-0.73) for the CuO$_2$ plane, (0.04,~-0.13,~0.05,~0.035,~-0.78) for the Hg-O state, and $t_12=0.01$ in eV.
\bibitem{RPAU}The components of spin interaction $\tilde{U}_s$ are defined as: intra-obital terms $(\tilde{U}_s)^{11}_{11}=U_1=1.74t_1$, $(\tilde{U}_s)^{22}_{22}=U_2=1.1t_1$ for the CuO$_2$ and Hg-O state, respectively and the inter-obital interaction between them $(\tilde{U}_s)^{12}_{12}=V=1.1t_1$. All values of $U$s are deduced by looking at their corresponding critical values above which the RPA denominator is positive and large, i.e. $(\tilde{U}_s)^{jj^{\prime}}_{ii^{\prime}} \le 1/(\tilde{\chi_0})^{jj^{\prime}}_{ii^{\prime}}$. 
\end{thebibliography}
\end{document}